# Applying different methods to model dry and wet spells at daily scale in a large range of rainfall regimes across Europe

Giorgio Baiamonte[1], Carmelo Agnese[1], Carmelo Cammalleri[2], Elvira Di Nardo[3], Stefano Ferraris[4], Tommaso Martini[3]

[1]Department of Agricultural, Food and Forest Sciences, University of Palermo, Viale delle Scienze 13, 90128 - Palermo, Italy
[2]Department of Civil and Environmental Engineering, Politecnico di Milano, Piazza Leonardo Da Vinci 32, 20133 – Milan, Italy
[3]Department of Mathematics "G. Peano", Università degli Studi di Torino, Via Carlo Alberto 10, 10123 - Torino, Italy
[4]Interuniversity Department of Regional and Urban Studies and Planning, Università degli Studi di Torino, Politecnico di Torino, Viale Pier Andrea Mattioli 39 10125 - Torino, Italy

*Correspondence to*: Giorgio Baiamonte (giorgio.baiamonte@unipa.it)

**Abstract.** The modelling of the occurrence of rainfall dry and wet spells (*ds* and *ws*, respectively) can be jointly conveyed using the inter-arrival times (*it*). While the modelling of *it* has the advantage of requiring a single fitting for the description of all rainfall time characteristics (including wet and dry chains, an extension of the concept of spells), the assumption on the independence and identical distribution of the renewal times *it* implicitly imposes a memoryless property on the derived *ws*, which may not be true in some cases. In this study, two different methods for the modelling of rainfall time characteristics at station scale have been applied: i) a direct method (DM) that fits the discrete Lerch distribution to *it* records, and then derives *ws* and *ds* (as well as the corresponding chains) from the *it* distribution; and ii) an indirect method (IM) that fits the Lerch distribution to the *ws* and *ds* records separately, relaxing the assumptions of the renewal process. The results of this application over six stations in Europe, characterized by a wide range of rainfall regimes, highlight how the geometric distribution does not always reasonably reproduce the *ws* frequencies, even when *it* are modelled by the Lerch distribution well. Improved performances are obtained with the IM, thanks to the relaxation of the assumption on the independence and identical distribution of the renewal times. A further improvement on the fittings is obtained when the datasets are separated into two periods, suggesting that the inferences may benefit for accounting for the local seasonality.

## 1 Introduction

Rainfall is an intermittent process that is characterized by the alternation of wet and dry statuses. Indeed, a very simple representation of the rain process consists in an alternating sequence of the two opposite conditions (dry/wet), each lasting for a given duration. Although some aspects of this intermittent process can only be observed at small time scales (i.e., hourly or smaller), particularly those concerning patterns of maximum intensity events, the daily time scale allows capturing the fundamental sequence of dry and wet events, as they are usually driven by the dynamic of large-scale precipitation systems (Bonsel and Lawford, 1999; Osei et al., 2021; Zhang et al., 2015). The daily scale is also quite appealing since precipitation records over several decades are reliably collected at this frequency, a feature seldom shared by sub-daily time-series.

Many approaches have been proposed in the scientific literature to model rainfall intermittence, including Poisson clusters, multifractals, power spectral analyses, Markov chains, and geo-statistics (Dey, 2023; Hershfield, 1970; Schleis, and Smith, 2016). At the local scale, a classical approach to address intermittency in rainfall records is to statistically analyse the sequences of rainy days, called wet spells and denoted as *ws*, and that of no-rainy days, called dry spells and denoted as *ds*, assumed to be independent of each other. When a dense network of stations is at hand, one way to properly account for the spatial correlation between stations is to consider Hidden Markov Models (Hughes and Guttorp, 1994; Robertson et al., 2004; 2008), to discover the existence of hidden weather states, or Markov Chains (Wilks, 1998) to model the rainfall amount process or occurrence process. In this paper, we are focusing on modelling rainfall intermittence locally, as a reasonable approach in case of data that are spatially independent.

In his pioneer study, Chatfield (1966) analysed a short series of daily rainfall recorded at a single station in Kew (London) and investigated the ratios between observed frequencies of increasing values of *ws*. Chatfield (1966) found that the probability that a wet day is followed by a rainy day is almost constant. This memoryless property characterizes the geometric distribution, and this assumption has been widely applied to describe the *ws* distribution in numerous studies since then (e.g., Kottegoda and Rosso, 1997; Racsko et al., 1991; Zolina et al., 2013). In the same study, Chatfield (1966) observed that "the probability that a dry day will be followed by a dry day does increase with the previous number of consecutive dry days", thus suggesting a tendency of the rainfall process to persist in the dry state. The Author proposed to adopt the log-series distribution to fit the *ds* series, as it exhibits increasing values of the subsequent ratios. While geometric and log-series distributions were applied in the past (Chatfield, 1966; Green, 1970) and are still often adopted to infer *ws* and *ds* probability laws (Chowdhury and Beecham, 2013; El Hafyani and El Himdi, 2022), some authors have questioned the general suitability of these distributions. As a few examples, Wilks (1999) suggested the use of a mixed geometric distribution for modelling *ws* over US, whereas Deni et al. (2010) assessed a good performance of the compound geometric distribution in Peninsular Malaysia. Furthermore, for both dry and wet spells, mixed distributions are generally observed to perform reasonably well (Dobi-Wantuch et al., 2000; Deni and Jemain, 2009).

Recently, Agnese et al. (2014) suggested to model both *ws* and *ds*, by investigating the probabilistic law of the so-called inter-arrival times (*it*), representing the series of times elapsed between two subsequent rainy days. By assuming that *it* are independent and identically distributed (i.i.d.) discrete random variables, any *it* series was interpreted as a realization of a sequence of holding times in a discrete renewal process. An important feature of any renewal process is that it restarts at each epoch of arrivals (the so-called *renewal property*).

Agnese et al. (2014) showed that both the *ws* and *ds* distributions can be easily derived from the *it* distribution. Indeed, the geometric distribution for *ws* directly arises from the i.i.d. hypothesis on the renewal times *it*, whereas the *ds* distribution follows the same probabilistic law adopted for fitting the *it* probabilities. This approach has been successfully applied over rainfall records collected in two different rainfall regimes in southern and northern Italy (Agnese at al., 2014; Baiamonte et al., 2019, respectively), and the 3-parameter Lerch-series distribution (part of the Hurwitz-Lerch-Zeta distributions family) has proven to reliably fit the *it* empirical frequencies, outperforming both the polylogarithmic and the logarithmic distributions (which are the discrete counterparts of two commonly adopted continuous distributions for dry spells).

Although the modelling of *ws* and *ds* distributions separately may allow a relaxation of i.i.d. hypothesis on *it*, modelling first the *it* sample has the clear advantage of retrieving the probability distribution of both *ws* and *ds* from a single fitting, often reducing the number of parameters. However, if the geometric distribution does not fit *ws* correctly, implying that the rainfall probability varies within the event, two separated models of *ws* and *ds* may be needed for a reliable evaluation of both quantities.

The objective of this paper is to verify to which extent the assumption of i.i.d. on renewal times *it* affects the ability to reproduce the probabilistic law of both *ws* and *ds*. The analysis is performed for six 70-year time-series recorded in sites with very different rainfall regimes across Europe (from southern Italy to northern Scotland). The appropriateness of the renewal property for the different sites is tested by comparing the outcomes of two modelling approaches: i) a direct method (DM), where the fitting is performed directly on *it*, whereas both *ws* and *ds* are simply derived from the previous *it* fitting, and ii) an indirect method (IM), where the renewal property is relaxed, and *ws* and *ds* are modelled separately, hence including the option to account for a not constant rain probability inside *ws*. The analysis is also extended to two additional time variables, strongly associated to *ws* and *ds*, called wet chains (*wch*, as previously introduced by Baiamonte et al., 2019) and dry chains (*dch*). These variables extend the concept of wet and dry spells to sequences characterised by an interruption of one no-rainy or one rainy day, respectively, as they represent two quantities that may be of interest for practical hydrological applications. Due to the likely influence of rainfall regimes on the results, the two methods (DM and IM) are evaluated not only on the entire time-series but also on two subsets obtained by following the rainfall seasonality in the study sites, as is common practice in ecohydrological applications.

## 2 Methods

### 2.1 The HLZ family of discrete probability distributions

The Hurwitz-Lerch Zeta (HLZ, also known as Lerch) family is a set of discrete probability distributions, whose probability mass function (*pmf*) is expressed in its general 3-parameter form as (Zörnig and Altmann, 1995):

$$p_v(k) = \frac{\theta^{k-1}}{(k+a)^s \Phi(\theta,s,a+1)} \qquad k = 1,2,\ldots \qquad (1)$$

where $\theta \in (0,1)$, $a > -1$, $s \in R$ are the parameters of the Lerch distribution, and the transcendent Lerch function $\Phi$ reads:

$$\Phi(\theta, s, a+1) = \sum_{n=0}^{\infty} \frac{\theta^n}{(n+a+1)^s}. \qquad (2)$$

Depending on the values assumed by the three parameters {$\theta, s, a$}, some special cases of the Lerch distribution family can be obtained, as summarized in **Table 1**. There, we reported only those cases with finite moments up to the order considered in this study. Note that non-negative values of *s* provide a monotonically decreasing *pmf* with mode equal to one.

This distribution makes it possible to account for some peculiar characters observed in the *it* samples, such as high values of both standard deviation and skewness, a monotonically decreasing pattern of frequencies with a slowly decaying tail, and a typical "drop" entering the dry status (*it*=2) from the wet status (*it*=1).

**Table 1 – Lerch family of probability distributions with the corresponding parameter domains, and the theoretical means (arithmetic, AM, geometric, GM, and/or harmonic, HM) that match (x) or not (–) with the empirical ones, according to the MLE method.**

| ID | Probability distribution | $\theta$ | s | a | AM | GM | HM |
|---|---|---|---|---|---|---|---|
| 1 | 3-par Lerch | 0<$\theta$<1 | > 0* | > -1 | x | x | x |
| 2 | 2-par polylogarithmic | 0<$\theta$<1 | > 0* | 0 | x | x | – |
| 3 | 1-par logarithmic | 0<$\theta$<1 | 1 | 0 | x | – | – |
| 4 | 1-par geometric | 0<$\theta$<1 | 0 | 1 | x | – | – |
| 5 | 2-par extended log | 0<$\theta$<1 | 1 | > -1 | x | – | x |

*The condition $s > 0$ allows obtaining monotonically decreasing *pmf*, with mode equal to 1.

To estimate the parameters of the Lerch distribution, the maximum likelihood estimation (MLE) method was applied. For the general case (3-parameter, Eq. 1), the analytical solution of the MLE method returns arithmetic, geometric and harmonic expectations equal to the corresponding sample values. If *N* denotes the sample size, the MLE function reads (Gupta et al., 2008):

$$ln\, L(\theta, s, a) = ln\, \theta \sum_{i=1}^{N} v_i - s \sum_{i=1}^{N} ln\, (a + v_i) - N\, ln\, [\Phi(\theta, s, a)] . \qquad (3)$$

For each special case of the family of Lerch distributions, **Table 1** also reports the constraints that need to be satisfied in the MLE. The likelihood equations were solved for both DM and IM by standard numerical methods to obtain the MLE.

While the general 3-parameter form (ID = 1 in **Table 1**) fits at least as well as any special cases having a smaller number of parameters (ID = 2–5 in **Table 1**), the inclusion of additional parameters is not always statistically justified (Wilks,1938). For this reason, the likelihood-ratio (LLR) test was employed to detect if the improvement in terms of log-likelihood is worth the introduction of the extra parameters. In this test, the *D* statistic was computed from the log-likelihood of the null model (ln($L_{ID}$) with ID = 2-5) and of the alternative model (ln($L_1$)) as:

$$D = -2[ln\, (L_{ID}) - ln\, (L_1)] \qquad (4)$$

where both log-likelihood values are computed from Eq. (3) using the corresponding parameterization. The $D$ values are approximately $\chi^2$ distributed, with degrees of freedom equal to the difference between the number of free parameters of the alternative and the null models.

To assess the adequacy of the Lerch family distribution in reproducing the observed frequencies, we employed a simulated $\chi^2$ goodness-of-fit test. When the distributions exhibit a long tail, the classical $\chi^2$ test might be biased due to the presence of numerous small class sizes (with less than 5 elements) and strong asymmetry (Martínez-Rodríguez et al., 2011). Therefore, we proceeded by reconstructing the distribution of the $\chi^2$ statistic under the null hypothesis via Monte Carlo simulation (Hope, 1968). Let us remind that this well-known goodness of fit test is characterized by the null hypothesis that the data follows the specified distribution, and by the alternative hypothesis that the data does not follow the specified distribution.

We simulated 3000 replicates of the sample by random sampling from the inferred theoretical distribution. The associated p-value (at a significance level of 0.05) was computed as the fraction of the 3000 replicates for which the computed $\chi^2_j$ values (with $j = 1, 2, …, 3000$) were greater than $\chi^2_{ref}$, that is the $\chi^2$ of the observed frequencies.

## 2.2 Inference of the main time properties of rainfall series

Let a time series of rainfall data be defined as H = $\{h_1, h_2, ..., h_n\}$, where $h$ (mm) is the rainfall depth recorded at a fixed uniform unit τ of time (e.g., a day). A day $k$ is considered rainy if the rainfall depth $h_k \geq h_*$, where $h_*$ is a fixed rainfall threshold. Thus, a sub-series of rainfall depth $h$ is selected from H, corresponding to the times E = $\{t_1, ..., t_{n_r}\}$ such that $h_{t_k} \geq h_*$ with $k \in \{1, ..., n_r\}$, and $n_r < n$ is an integer multiple of the timescale τ.

The inter-arrival time series $it = \{it_1, it_2, ..., it_{nr}\}$ is defined as the sequence built with the times elapsed between each element of E (except the first one) and the immediately preceding one. It is worth noting that if $it_k = 1$, the rainy day $k$ immediately follows the rainy day $k-1$; consequently, in the $it$ time-series, any sequence of $m$ consecutive unitary values is a wet spell (*ws*) of duration $m+1$. Keeping in mind this observation on *ws*, the isolated rainy day $k$ is identified if the following condition is satisfied:

$$it_k > 1 \text{ and } it_{k+1} > 1. \tag{5}$$

Additionally, the sequence of dry spells (*ds*) can be derived from the *it* sequence as:

$$ds_k = it_k - 1 \text{ for any } it_k > 1. \tag{6}$$

We define "wet chain", *wch*, the sequence of wet spells broken by 1-day dry spells. This concept was introduced in Baiamonte et al. (2019), namely 'wet day sequences', and it corresponds to a *ws* if no interruption occurs, whereas, if an event $ds > 1$ occurs, the wet chain runs out. Similarly, we introduce here "dry chain", *dch*, a sequence of dry spells interrupted by 1-day wet spells. A *dch* corresponds to a *ds* if no interruption occurs, whereas a *dch* expires if a *ws* > 1 occurs. Different examples of *wch* and *dch* are given in **Fig. 1**.

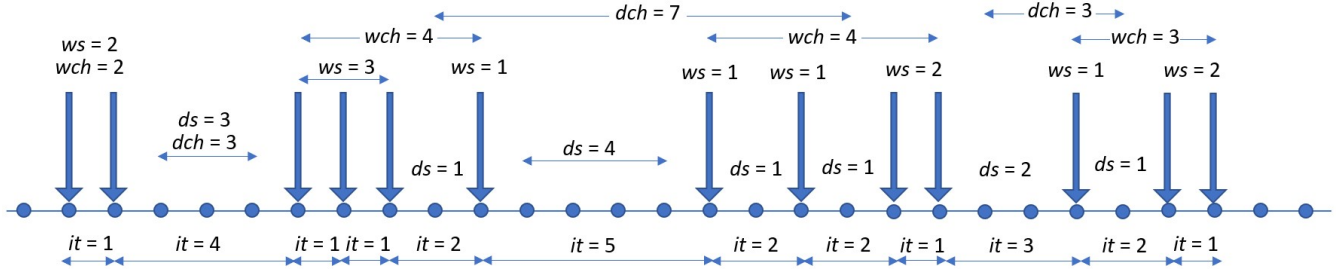

**Figure 1:** Examples of interarrival times (*it*), wet spells (*ws*), dry spells (*ds*), wet chains (*wch*), and dry chains (*dch*).

From a hydrological point of view, the above-described *wch* can be seen as a single rainy period in a broad sense, since a single no-rainy day in between multiple rainy days does not significantly alter the wet status, as if the entire period is likely related to the same meteorological perturbation. A similar consideration can be made for *dch*, since a single rainy day that interrupts a sequence of dry spells also may not significantly alter the dry status, with the additional caveat that the rainfall depth during the isolated rainy day should be of limited amount. Following the consideration that an isolated rainy day (surrounded by two dry spells) is likely associated to only few hours of rain, the assumption of limited rainfall depth seems reasonable when only durations are analyzed.

### 2.2.1 The direct method

According to previous studies (Agnese at al., 2014; Baiamonte et al., 2019), a direct inference on the *it pmf* (from now on referred to direct method, DM) can be performed by Eq. (1) (with *v* = *it*). Following the i.i.d. hypotheses on *it*, the random variable *ws* results geometrically distributed, with parameter $(1 - p_{it}(1))$, where $p_{it}(1)$ is the probability that *it* is equal to 1 as given in Eq. (1). Thus, we have:

$$p_{ws}(k) = \big(1 - p_{it}(1)\big) p_{it}(1)^{k-1} \qquad k = 1,2,\ldots \qquad (7)$$

where $p_{it}(1)$ is equal to:

$$p_{it}(1) = \left(\frac{1}{(a+1)^s \Phi(\theta,s,a+1)}\right). \qquad (8)$$

From Eq. (7), the *ws pmf* satisfies the memoryless property of the geometric distribution, that is:

$$P(ws > k + m | ws > m) = P(ws > k). \qquad (9)$$

Eq. (9) is the consequence of the simple but restrictive hypothesis that the probability of future rainy day occurrence is not affected by the occurrence of past rainy days, i.e. the probability of a rainy day is assumed to be constant (Chatfield, 1966).

According to Eq. (6), the *ds pmf* can be easily recovered from the *it pmf* as (Agnese et al., 2014):

$$p_{ds}(k) = \frac{p_{it}(k+1)}{1 - p_{it}(1)}. \qquad (10)$$

The *wch pmf* follows from the *it* distribution too, by considering that a wet chain is a sequence of *it* = 1 (consecutive rainy days) and *it* = 2 (corresponding to a hole in between two rainy days), ending with *it* > 2. Therefore, taking into account the probabilities that *it* is equal to either 1 or 2, as well as all the probability of different combinations of *it* values, we can write:

$$p_{wch}(k) = (1 - p_{it}(1) - p_{it}(2)) \sum_{j=0}^{k-1} \binom{k-1}{j} p_{it}(1)^{k-1-j} p_{it}^{j}(2) \qquad (11)$$

where *j* (0 ≤ *j* < *k*) is the number of holes breaking the wet status, $\binom{k-1}{j}$ is the number of subsets of *j* elements chosen among *k*–1 elements, and *k*–1–*j* is the number of *it*=1, indicating a wet spell of size *k*–1. It is worth highlighting that $p_{wch}(1) = (1 - p_{it}(1) - p_{it}(2)) = P(it > 2)$.

An alternative, and more general, formulation of the *wch pmf*, $p_{wch}(m)$, relies on a convolution approach. In this case, $p_{wch}(m)$ can be obtained following two observations: i) that the probability of the sum of *k* wet spells of any duration ($P(\sum_{j=1}^{k} ws_j = m)$, with *k* = 1,2,…) is equal to the *k*-fold convolution of *ws* with itself, corresponding to $p_{ws}^{k*}(m)$, and ii) that the probability to have a sequence of *k*–1 1-day dry spells, corresponding to $P(ds^{(1)}=1, \ldots ds^{(k-1)}=1)$, can be expressed as $p_{ds}(1)^{k-1}(1 - p_{ds}(1))$. By extending the convolution over all the possible combinations of *ws* sequences, summing to *m* and to the *n* 1-day *ds* sequences (0 ≤ *n* ≤ *m*-1) breaking the wet status, the probability $p_{wch}(m)$ can be expressed as:

$$p_{wch}(m) = \sum_{k=1}^{m} p_{ds}(1)^{k-1}(1 - p_{ds}(1)) p_{ws}^{k*}(m). \qquad (12)$$

Eq. (11) follows from Eq. (12) when the hypothesis of i.i.d. on renewal times *it* holds.

Using similar arguments but with *ds* in place of *ws*, the *dch pmf*, $p_{dch}(m)$, results to be:

$$p_{dc}(m) = \sum_{k=1}^{m} p_{ws}(1)^{k-1}(1 - p_{ws}(1)) p_{ds}^{k*}(m), \qquad (13)$$

which can be obtained from Eq. (12) by simply substituting *ws* by *ds* and vice versa. Note that $p_{dch}(1) = p_{ds}(1)(1- p_{ws}(1))$. It should be noted that the definition of chains can be easily extended to interruptions longer than a single day (i.e., a 2-day *wch* is a sequence of *it* = 1, *it* = 2, and *it* = 3, ending with *it* > 3). However, the chains can become less and less realistic with increasing length of the interruption.

Equations (7), (10), (11) or (12), and (13) show that in the DM the probability distributions of the wet and dry spells, and of the wet and dry chains, can be directly recovered from the *it* distribution, without the need of extra fittings or parameterizations.

### 2.2.2 The indirect method

The indirect method (IM) is based on modelling individually the random variables *ws* and *ds*, and it is the most commonly adopted approach for the modelling of *ws* and *ds* in hydrology (Kottegoda and Rosso, 2008). Both distributions rely on Eq. (1) (*v* = *ws* and *v* = *ds*). In this case, the observations of *ws* and *ds* are assumed to be i.i.d. and independent of each other. Note that, contrary to what has been observed in many studies and described in section 2.2.1, in the IM the *ws* distribution is not constrained to be geometric, since the hypothesis of constant probability - when a wet day is followed by another wet day - is removed. Indeed, we have:

$$\frac{P(ws>r+1)}{P(ws>r)} = 1 - \frac{P(ws=r+1)}{P(ws>r)}, \tag{14}$$

where the left-hand term is the so-called failure rate, $FR_{ws}(r)$. If the *ws* pmf is assumed to belong to the family of Lerch distributions, the failure rate reads (Gupta et al., 2008):

$$FR_{ws}(r) = \frac{1}{(a+r)^s \Phi(\theta,s,a+r)}. \tag{15}$$

If we set *s* = 0, which corresponds to $Geom(1-\theta)$ in the family of Lerch distributions (**Table 1**), we get $FR_{ws}(r) = \theta$ and thus $\frac{P(ws>r+1)}{P(ws>r)} = 1 - \theta$ is constant, as expected. For all the other distributions in the Lerch family, $FR_{ws}(r)$ will suitably depend on *r* allowing for more flexibility (Gupta et al., 2008).

Once both the *ws* and *ds* distributions are known, the *it* distribution can be easily recovered as:

$$p_{it}(1) = \frac{\sum_{j=1}^{\infty}(j-1)\,p_{ws}(j)}{\sum_{j=1}^{\infty} k\,p_{ws}(j)} = \frac{E[ws]-1}{E[ws]} \qquad \text{if } k = 1, \tag{16}$$

$$p_{it}(k) = p_{ds}(k-1)(1-p_{it}(1)) \qquad \text{if } k > 1. \tag{17}$$

The procedure to derive *it* from *ws* and *ds* is not commonly discussed in studies adopting the IM method, as they are usually focused only on the modelling of *ws* and *ds* rather than *it*.

As in the DM, the probability distribution of *wch* and *dch* can be obtained from Eq. (12) and Eq. (13), respectively, where the probabilities $p_{ds}$ and $p_{ws}$ are now separately inferred, unlike in the DM for which these probabilities are recovered from the *it* distribution. It is worth mentioning that for the *wch pmf*, Eq. (11) is no longer valid, since the assumption of memoryless property is relaxed. From a computational point of view, we emphasize that the IM will always require two fittings (*ws* and *ds*) performed through a numerical MLE, unlike the DM where only one fitting is sufficient to recover the distribution of *it*.

## 3 Materials

### 3.1 Rainfall records

In this paper, to obtain the series of rainfall time variables (*it*, *ws*, *ds*, *wch*, *dch*), daily rainfall records are analysed. To discriminate between rainy and no-rainy days, a fixed rainfall threshold $h^* = 1$ mm is chosen, according to the conventional value established by the World Meteorological Organization. The dataset includes six time series collected over Europe at different latitudes (from 38° N to 58° N), from Trapani and Floresta in Sicily to Stornoway in Northern Scotland. **Figure 2** depicts the spatial location and altitude of the six stations. About 70 years of recorded data are used for each rain gauge: Ceva (1950-2016), Floresta (1951-2015), Oxford (1950-2017), Stornoway (1950-2020), Torino (1950-2017), and Trapani (1950-2015), with a minimal occurrence of missing data.

Due to the relevance of rainfall regimes on the distribution or rainy and no-rainy days, in addition to the analysis of the entire year (Y), all analyses were performed for two additional data subsets: (S1) from April to September, and (S2) from October to March.

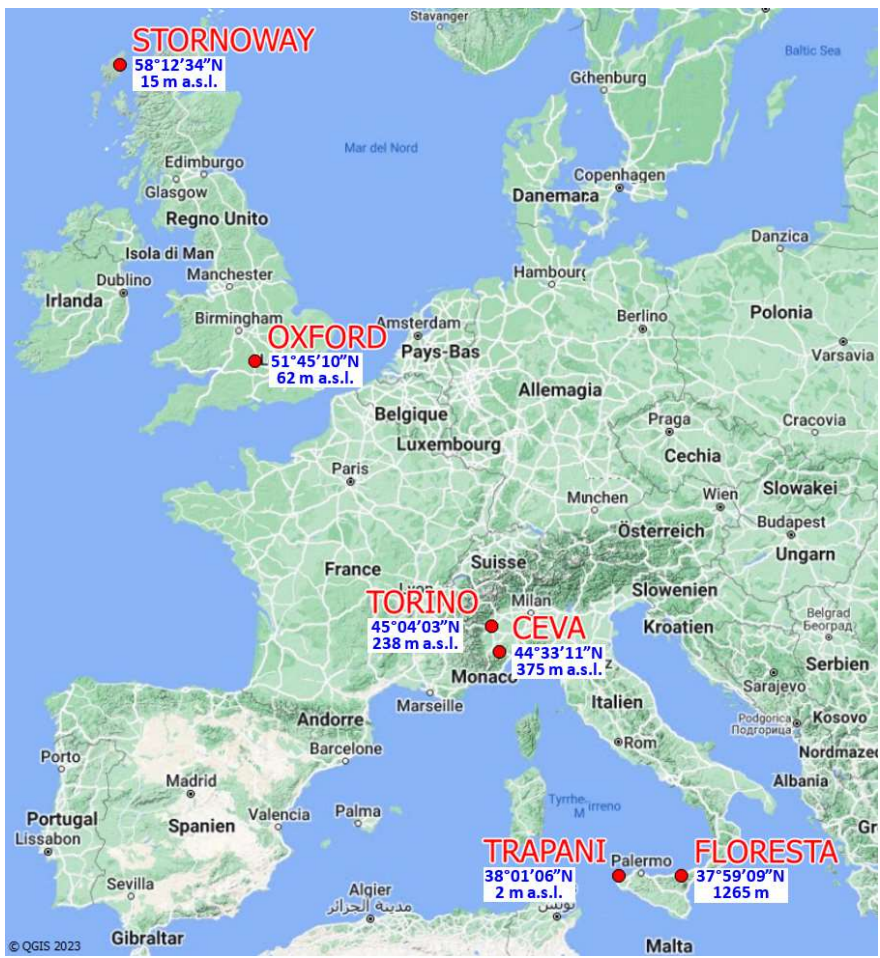

**Figure 2**: Locations of the six considered stations, together with latitudes and elevations above the sea level (© QGIS 2023).

The analysed stations are characterized by different rainfall regimes, as shown in **Fig. 3** by the average number of rainy days in each month (panel a), and the fraction of the yearly-average rainy days in each month (panel b) (as a standardization of the data in panel a). The stations of Trapani (TRA) and Floresta (FLO) represent the typical Mediterranean climate, with a strong seasonality in the rainfall regime (see **Fig. 3b**), and precipitations concentrated in the S2 season. The two stations differ in the total amount of average annual rainfall, which is low for TRA (420 mm) and high for FLO (1133 mm), a difference that is also revealed by the different number of rainy days per month (**Fig. 3a**). Torino (TOR) and Ceva (CEV) are characterised by a mid-latitude sublitoranean climate, with a high rain frequency in spring (**Fig. 3b**). CEV also exhibits a secondary peak in autumn, mainly due to the influence of the Tyrrhenian Sea warming in summer. Despite this difference, the two stations are characterised by similar total annual rainfall (829 and 836 mm, respectively) and number of rainy days (**Fig. 3a**). Oxford (OXF) is a Northern Europe station with a relatively low average rainfall amount (592 mm) homogeneously distributed throughout the year, whereas Stornoway (STW) has very high rain frequency and a higher amount all over the year (1072 mm), due to its location in the far north-western Scotland and the direct effect of the wet fronts from the Atlantic Ocean. Both stations in the UK have low seasonality compared to the other stations (see **Fig. 3b**).

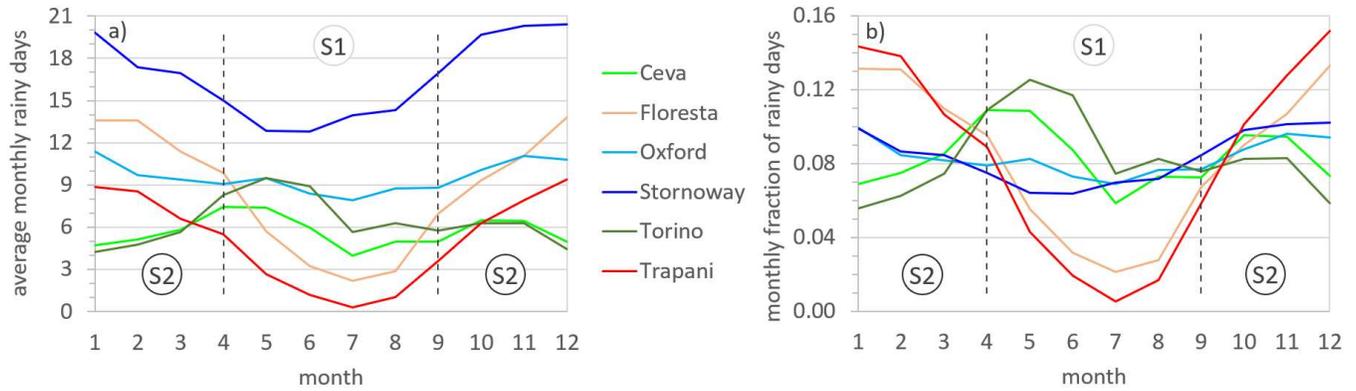

**Figure 3: Time variability of: a) average number of rainy days in each month, and b) fraction of yearly-average rainy days in each month. Dashed lines delimit the two seasons S1 (April – September) and S2 (October – March).**

For the stations of TRA and FLO, seasons S1 and S2 clearly correspond to the low and high frequency of rain events, respectively (**Fig. 3b**). A similar pattern can be observed for OXF and STW, although with less marked differences between the two seasons. Due to the considerable length of the data records, sample sizes remain large even for the two seasonal datasets, as summarised by the data in **Table 2**. The sample size is less than 500 only for *dch* in TRA for S1, which can be explained by the numerous long dry periods occurring in the dry season.

**Table 2 – Sample sizes of the time variables for the six stations and for the three periods, Year, S1 (from April to September), and S2 (from October to March).**

| Station | Year | | | | S1 | | | | S2 | | | |
|---|---|---|---|---|---|---|---|---|---|---|---|---|
| | it | ws – ds | wch | dch | It | ws – ds | wch | Dch | it | ws – ds | wch | dch |
| CEV | 4569 | 2530 | 2058 | 1102 | 2335 | 1357 | 1159 | 528 | 2203 | 1154 | 890 | 568 |
| FLO | 6495 | 2867 | 2048 | 1570 | 1966 | 1053 | 858 | 481 | 4520 | 1811 | 1188 | 1087 |
| OXF | 7933 | 3920 | 2559 | 1915 | 3664 | 1883 | 1254 | 902 | 4262 | 2034 | 1303 | 1012 |
| STW | 14227 | 4126 | 2202 | 2649 | 6065 | 2216 | 1259 | 1337 | 8145 | 1909 | 943 | 1312 |
| TOR | 5171 | 2726 | 2145 | 1289 | 3023 | 1668 | 1277 | 726 | 2147 | 1057 | 868 | 563 |
| TRA | 4064 | 2256 | 1755 | 970 | 955 | 661 | 574 | 220 | 3108 | 1594 | 1181 | 750 |

It is noteworthy that the splitting of the two seasons of CEV and TOR was done differently in a previous paper (Baiamonte et al., 2019). However, in this paper the same splitting in two six-month seasons is used for the sake of the homogeneity of the present analysis (**Fig. 3a and 3b**).

## 3.2 Preliminary tests on observed records

As is well known in the literature, the presence of a trend in the datasets can affect the assumptions made for *it* (in the DM) or for *ws* and *ds* (in the IM). Therefore, before fitting, a trend test was performed. We used the well-known Mann-Kendall (MK) non-parametric test (Mann, 1945; Gilbert, 1987) at a significance level of 0.05. Note that the MK test is frequently used in literature for detecting significant trends in hydro-meteorological time series (see for example Gocic and Trajkovic, 2013, and references therein).

However, a known limitation of the MK test is the increased probability of finding trends in the presence of a significant autocorrelation in the data (Hamed and Rao, 1998). In such a case, the variance of the MK test statistic depends on the true unknown autocorrelation structure, and it is typically larger (lower), if positive (negative) autocorrelation occurs with respect to the case of independent data. Therefore, in the presence of autocorrelation a correction is needed, as the critical values of the classical MK test would lead to incorrect results. Hamed and Rao (1998) proposed an approximation of the true variance of the MK test statistic in case of autocorrelated data.

Let us recall that a key difference between the DM and the IM, as introduced in section 2.2, is related to modelling the *it* samples as i.i.d. renewal times, which results in a geometric distribution of *ws* in the DM. One way of directly testing the memoryless property of the empirical data is to study the behaviour of the ratios $\frac{S_{r+1}}{S_r}$ for all $r \in \{1, ..., \max(ws)-1\}$, with $S_r$ the number of wet spells of length at least *r* in a *ws* series. The memoryless property implies that these ratios are constant, while they would depend on *r* for a *pmf* in the general form of the Lerch distribution (Gupta et al., 2008).

## 4 Results

### 4.1 Non-parametric analyses

The presence of a trend on the recorded *it*, *ws* and *ds* time-series was tested using the Kendall's τ, which is a non-parametric measure of rank correlation. The standard MK tests - applied using the τ values reported in **Table 3** - shows the presence of a small statistically significant trend in some cases (indicated in bold in the table). Since a weak positive autocorrelation was always observed (**Table 3**) for the three variables *it*, *ws* and *ds* in each of the six stations, the corrected MK was implemented through the Fume package of the R software (2019). This corrected test returned the absence of a statistically significant trend for all the variables and stations; therefore, the entire dataset can be adopted for subsequent analyses.

Table 3 – Values of Kendall's τ statistic for Year period time-series of the six stations. In bold, the values that are statistically significant (p = 0.05) for the classical MK test but not for the corrected MK test. All the other values were found to be not significant even with the classical MK test.

| Station | it | ws | ds |
|---|---|---|---|
| CEV | **0.0261** | -0.0237 | **0.0309** |
| FLO | **-0.0201** | 0.0244 | -0.0040 |
| OXF | -0.0020 | -0.0002 | 0.0128 |
| STW | **-0.0217** | **0.0338** | -0.0048 |
| TOR | 0.0049 | 0.0180 | **0.0329** |
| TRA | 0.0181 | -0.0034 | 0.0204 |

To verify the memoryless property of *ws* directly on the empirical data, **Fig. 4** plots the sequence of the ratio $\frac{S_{r+1}}{S_r}$ for all stations and periods. The series are reported up to *ws* values with a number of observations greater or equal to 10. These results show a roughly constant value for the two stations of CEV and TOR (**Fig. 4a**), a slightly increasing trend for TRA and FLO (**Fig. 4b**) and a marked increasing trend for OXF (**Fig. 4c**) and STW (**Fig. 4d**). These results suggest that the use of a geometric distribution for the *ws* records of OXF and STW may not be adequate, as successively investigated with the parametric analysis, since the variability of the $\frac{S_{r+1}}{S_r}$ ratios could be considered as a rough indicator of the inadequacy of the geometric distribution in describing wet spells (Chatfield, 1966).

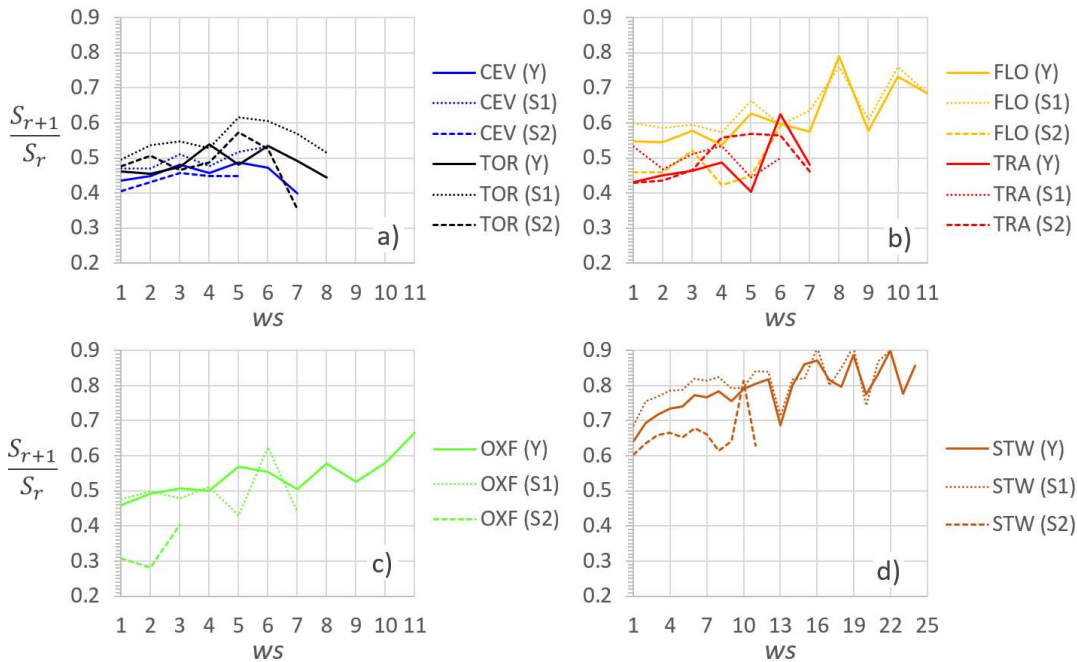

**Figure 4: For the three periods, Y, S1 and S2, $S_{r+1}/S_r$ ratio versus *ws*, a) for CEV and TOR, b) for FLO and TRA, c) for OXF, and d) for STW.**

The main statistics of all rainfall time variables, *it*, *ws*, *ds*, *wch* and *dch*, are summarized in the box plots depicted in **Fig. 5**, reporting the statistics for all the stations and for the three periods, Y, S1 and S2. The figure highlights the different statistical characters of the investigated variables, and how the seasonality seems to affect all sites.

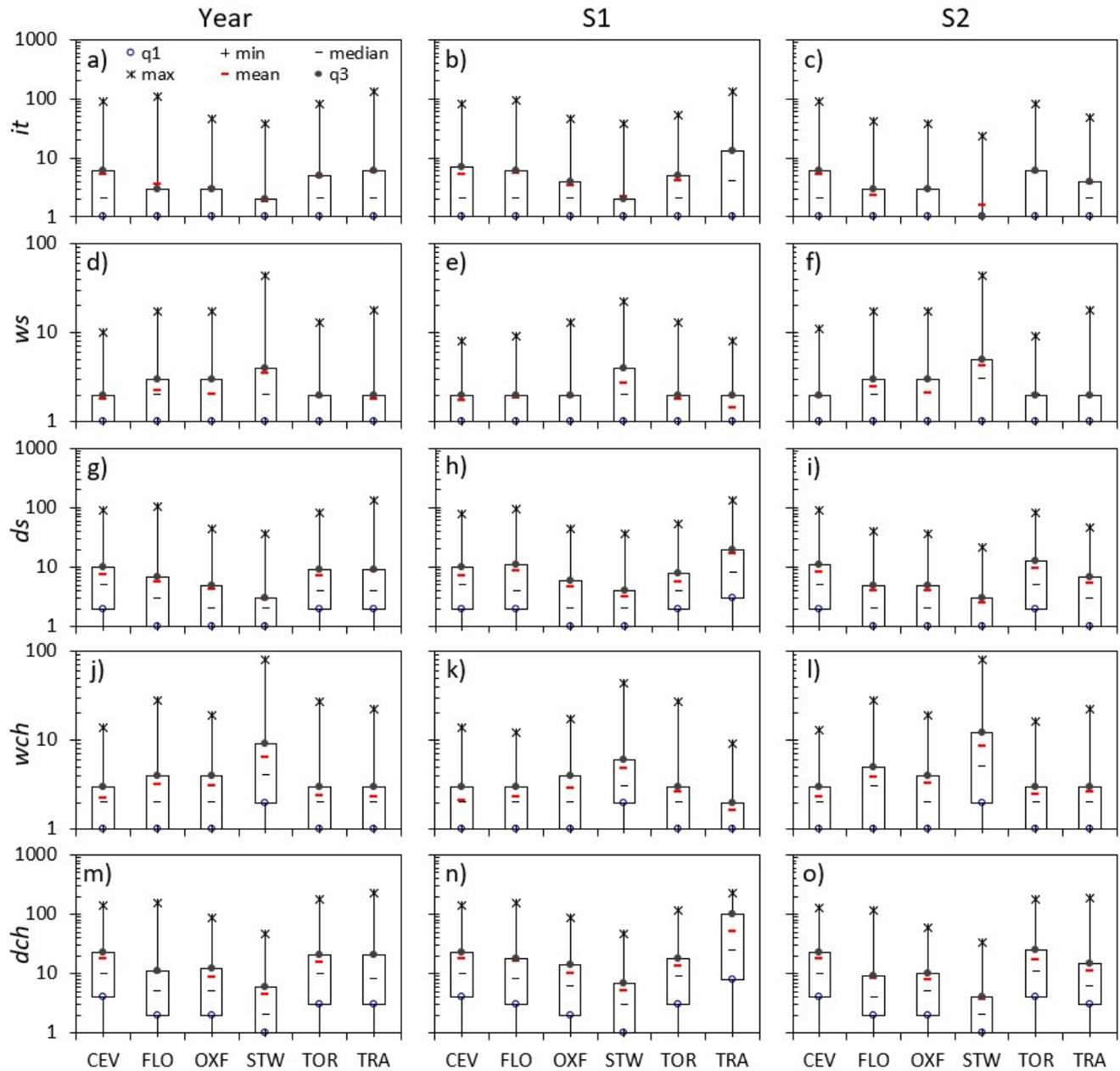

**Figure 5:** Box plots of the time variables, *it* (a-c), *ws* (d-f), *ds* (g-i), *wch* (j-l) and *dch* (m-o) for all six stations and for the three periods, year (left column), S1 (central column) and S2 (right column). The variables q1 and q3 identify the first and the third quartiles, respectively.

It is interesting to observe that the STW station shows the highest *ws* and *wch* statistics, likely due to its high frequency rainfall. These high values are obviously counterbalanced by the lowest *it*, *ds* and *dch* statistics for all the considered periods. For *ws* (S1), almost all the stations show a quite limited range of variability, with the exception of STW and TRA that present very different rainfall regimes.

In depth analysis of the relationship between spells and chains can be made on the data reported in **Fig. 6** where, for the six stations, and for the two seasons S1 (a,c) and S2 (b,d), the ratios of the observed cumulated frequencies $F_{ws}/F_{wch}$ (a,b) and $F_{ds}/F_{dch}$ (c,d) versus the corresponding time variables are plotted. These ratios describe the relative weight of the derived variable *dch* (*wch*) on *ds* (*ws*). As expected, the ratios $F_{ws}/F_{wch}$ and $F_{ds}/F_{dch}$ are greater than the unity, with larger values for $F_{ds}/F_{dch}$ [1-3] than for $F_{ws}/F_{wch}$ [1-1.7]. In general, moving from S1 to S2, higher $F_{ws}/F_{wch}$ and lower $F_{ds}/F_{dch}$ values can be observed. Ratios corresponding to CEV and TOR rain gauges provide almost similar values for the two seasons, according to the limited seasonality, whereas STW, characterized by a very high rainfall frequency, is the only case with a ratio $F_{ws}/F_{wch}$ higher than $F_{ds}/F_{dch}$. Finally, it is worth noticing the very high values of $F_{ds}/F_{dch}$ for TRA in the S1 period, reflecting the aridity that characterizes the area of western Sicily during that season, and the large differences in $F_{ws}/F_{wch}$ in FLO when comparing the two seasons.

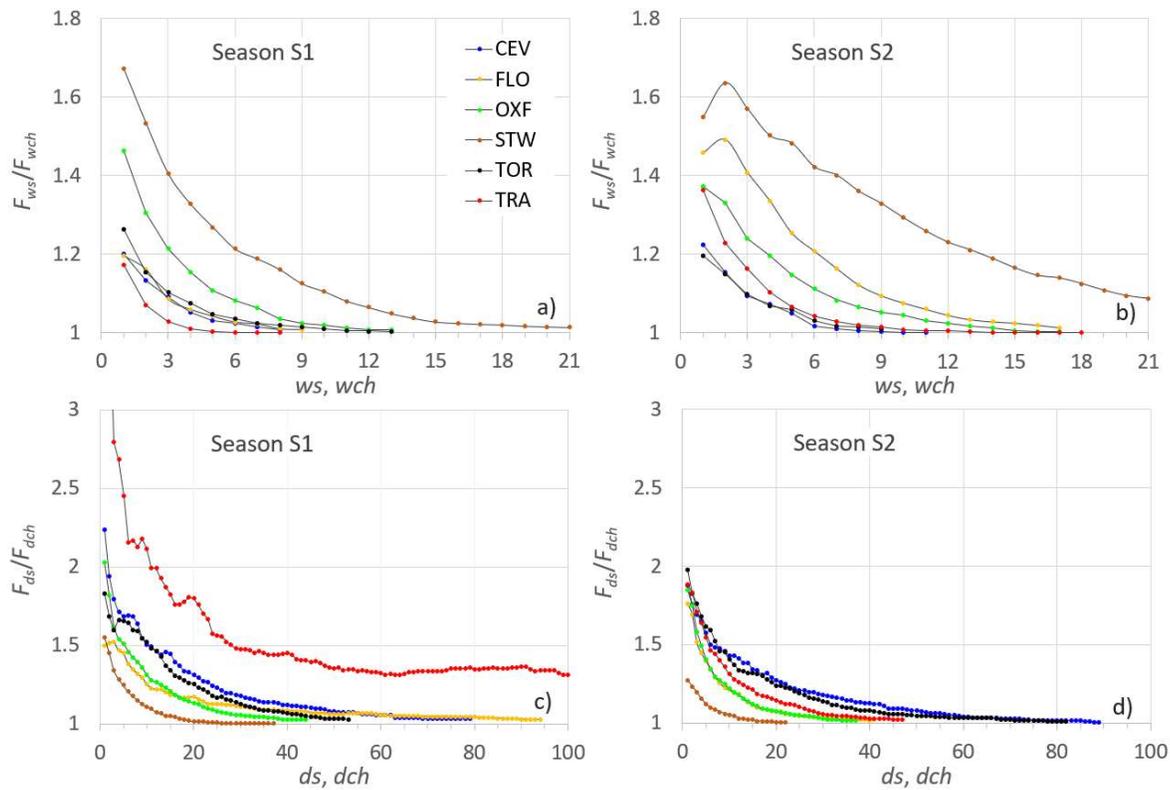

**Figure 6:** Ratios between observed cumulative frequencies $F_{ws}/F_{wch}$ (panels a,b) and $F_{ds}/F_{dch}$ (panels c,d) for the six stations, and for S1 (panels a,c) and S2 (panels b,d).

## 4.2 DM and IM comparison

For the time-series of the six rain gauges, the Lerch family (as given in Eq. (1)) was fitted on the three periods (Y, S1 and S2), for *it* (DM) and for both *ws* and *ds* (IM). The parameters obtained by MLE are summarised in **Table 4**. It is important to highlight that, for each combination station/period, the parameters reported in **Table 4** correspond to the Lerch family special cases (see **Table 1**), as supported by the results of the LLR test (e.g., 3-parameters are adopted only when this number of parameters is justified by the test).

**Table 4 – Parameters of the Lerch family of probability distributions fitted on *it* (DM), *ws* and *ds* (IM) for the six stations and for the three periods, Year, S1 and S2. Values of 0 or 1 for the parameters *s* and *a* identify the special cases listed in Table 1.**

| Station | variable | $\theta$ | | | $s$ | | | $a$ | | |
|---|---|---|---|---|---|---|---|---|---|---|
| | | Y | S1 | S2 | Y | S1 | S2 | Y | S1 | S2 |
| CEV | | 0.913 | 0.902 | 0.923 | 0.442 | 0.395 | 0.476 | -0.953 | -0.954 | -0.958 |
| FLO | | 0.934 | 0.942 | 0.858 | 1.005 | 0.700 | 0.738 | -0.657 | -0.833 | -0.809 |
| OXF | *it* | 0.902 | 0.906 | 0.896 | 1.069 | 0.996 | 1.122 | -0.384 | -0.437 | -0.342 |
| STW | | 0.867 | 0.864 | 0.861 | 1.348 | 1.186 | 1.539 | -0.528 | -0.489 | -0.508 |
| TOR | | 0.919 | 0.871 | 0.940 | 0.581 | 0.393 | 0.549 | -0.891 | -0.942 | -0.953 |
| TRA | | 0.970 | 0.975 | 0.897 | 1.164 | 0.744 | 0.689 | -0.364 | -0.687 | -0.771 |
| CEV | | 0.446 | 0.419 | 0.476 | 0 | 0 | 0 | 1 | 1 | 1 |
| FLO | | 0.650 | 0.464 | 0.599 | 1 | 0 | 0 | 3.084 | 1 | 1 |
| OXF | *ws* | 0.558 | 0.486 | 0.600 | 0.261 | 0 | 0.382 | 0 | 1 | 0 |
| STW | | 0.843 | 0.696 | 0.853 | 1 | 0.324 | 0.585 | 0.921 | 0 | 0 |
| TOR | | 0.473 | 0.583 | 0.508 | 0 | 1 | 0 | 1 | 1.143 | 1 |
| TRA | | 0.553 | 0.308 | 0.486 | 1 | 0 | 0 | 1.896 | 1 | 1 |
| CEV | | 0.913 | 0.901 | 0.922 | 0.433 | 0.387 | 0.467 | 0 | 0 | 0 |
| FLO | | 0.967 | 0.953 | 0.853 | 1.938 | 1 | 0.662 | 2.889 | 1.399 | 0 |
| OXF | *ds* | 0.890 | 0.880 | 0.880 | 0.827 | 0.543 | 0.835 | 0 | -0.548 | 0 |
| STW | | 0.838 | 0.861 | 0.799 | 1 | 1 | 1 | 0 | 0 | 0 |
| TOR | | 0.918 | 0.871 | 0.940 | 0.552 | 0.384 | 0.541 | 0 | 0 | 0 |
| TRA | | 0.989 | 0.980 | 0.892 | 2.042 | 1 | 0.614 | 3.870 | 2.062 | 0 |

In the case of *it* (DM), the 3-parameter Lerch distribution is selected for all sites and periods; the $\theta$ parameter varies between 0.86 and 0.97, with higher values observed for the whole year and season S1 for TRA. The variation of $\theta$ in the various periods is very limited for OXF, STW, and to a lesser extent CEV, as evidenced by a low degree of seasonality compared to the other stations. As expected, the *a* values are all negative (**Table 4**), since the *a* parameter allows reproducing the observed drop in frequency when moving from *it* = 1 to *it* = 2. The *s* parameter is positive for all the *it* fittings, suggesting that the mode is always at *it* = 1.

For the IM, the geometric distribution (*s* = 0 and *a* = 1) was chosen in several cases (10 out of 18) to fit *ws*, most commonly over the sub-seasons (8 out of 12) than over the whole year. CEV station is the only one where the geometric distribution was selected for all three periods, while the geometric distribution was never selected for STW. To capture the probabilistic law of *ws*, OXF and STW seem to require the polylog-series or the 2-par extended log-series distribution (**Table 1**). It is interesting to observe that for TRA and FLO, the geometric distribution is not appropriate for *ws* when the period is the year, while it is selected when the complete dataset is split into two sub-seasons.

Concerning the *ds* distribution in the IM, two or three parameters are always required, with the notable exception of STW, where the logarithmic distribution ($s = 1$ and $a = 0$) is selected for all three periods. The polylog-series distribution is selected for 10 out of 18 cases, confirming the adequacy of this distribution to describe the probability law of dry spells in different cases; noteworthy, for CEV and TOR the values assumed by $\theta$ and $s$ in the polylog distribution of *ds* are almost equal to those of the Lerch distribution for *it*. Hence for these stations, the additional parameter *a* in the Lerch distribution fulfils, in practice, the function of accounting for the geometric distribution of *ws*.

The assessment of the goodness-of-fit for the selected distributions (for *it* in the DM, and for *ws* and *ds* in the IM) is synthetically illustrated in **Fig. 7**. In particular, for all stations and periods (Y, S1, S2) the computed p-values are classified according to four ranges (0-0.01, 0.01-0.05, 0.05-0.1, 0.1-1), with light (0.05-0.1) and dark (0.1-1) green classes referring to the acceptance range of the null hypothesis. For a few of the 180 (6 stations × 3 periods × 5 variables × 2 methods) Monte Carlo procedures (19/180), the presence of outliers (high values with very low frequency) suggested a preliminary data smoothing. Such smoothing was performed by uniformly distributing the frequency of the outlier over all values between the observed value and the latest observed non-null frequency. In **Fig. 7**, these cases are marked by black dots. Overall, **Fig. 7** shows that the fitting of *it* is satisfactory in all cases for Y, with only one exception (FLO) with a p-value just above 0.05. Analogously, good results are obtained for *ds*, which is not surprising given the close relation between *it* and *ds* modelling in the DM framework (see Eq. 6). In contrast, the fits are not satisfactory in several cases for the resulting *ws* and *wch*, most notably for STW and TRA.

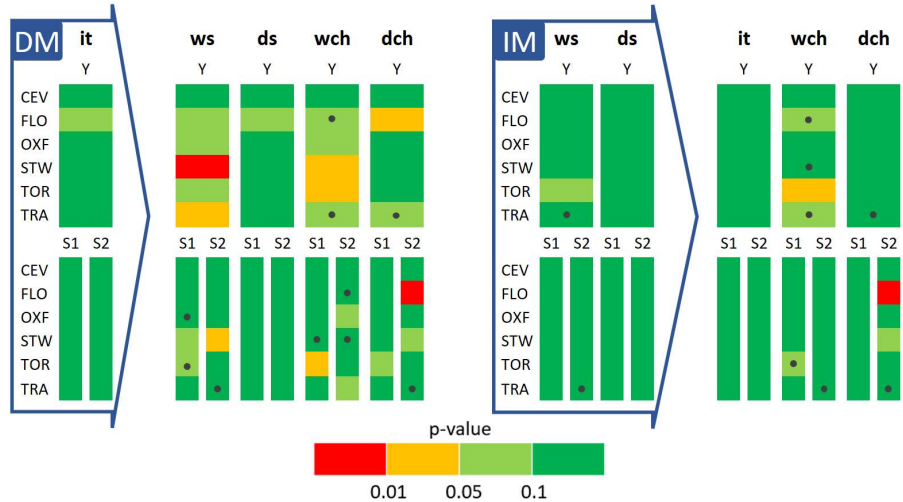

**Figure 7:** Summary of the results of the $\chi^2$ test for both the direct (DM, left-side figure) and the indirect (IM, right-side figure) methods. The variables inside the large arrows are the ones fitted in the corresponding method, whereas the other variables are deducted. The p-values (see legend) for all the stations and periods are reported. Black dots indicate that smoothing of observed frequencies was applied to calculate $\chi^2_{ref}$.

The data depicted on the right side of **Fig. 7** (i.e., IM) suggests that when the IM is applied, there is a significant reduction in the number of unsatisfactory fits (red and orange classes), particularly for *ws* and *wch*. In addition, when analysing the seasonal datasets, a further improvement in the identification of the probability law of the time variables is evident.

The plots in **Fig. 8** and **9** show the cumulative observed frequencies and the corresponding fitted Lerch family probability distributions for the annual period (Y) when applying the DM and the IM, respectively. The comparison between the two methods confirms the overall net improvement achieved by the IM in those cases that were not well fitted in the DM, such as *ws* and *wch* for STW (compare **Fig. 8b** and **8d** with **Fig. 9a** and **9c**). The latter consideration is also supported by the comparison between the p-values classes (**Fig. 7**) associated with the same variables, when IM is used in place of DM. This result underlines how the i.i.d. hypothesis for *it* in the DM may not always be valid.

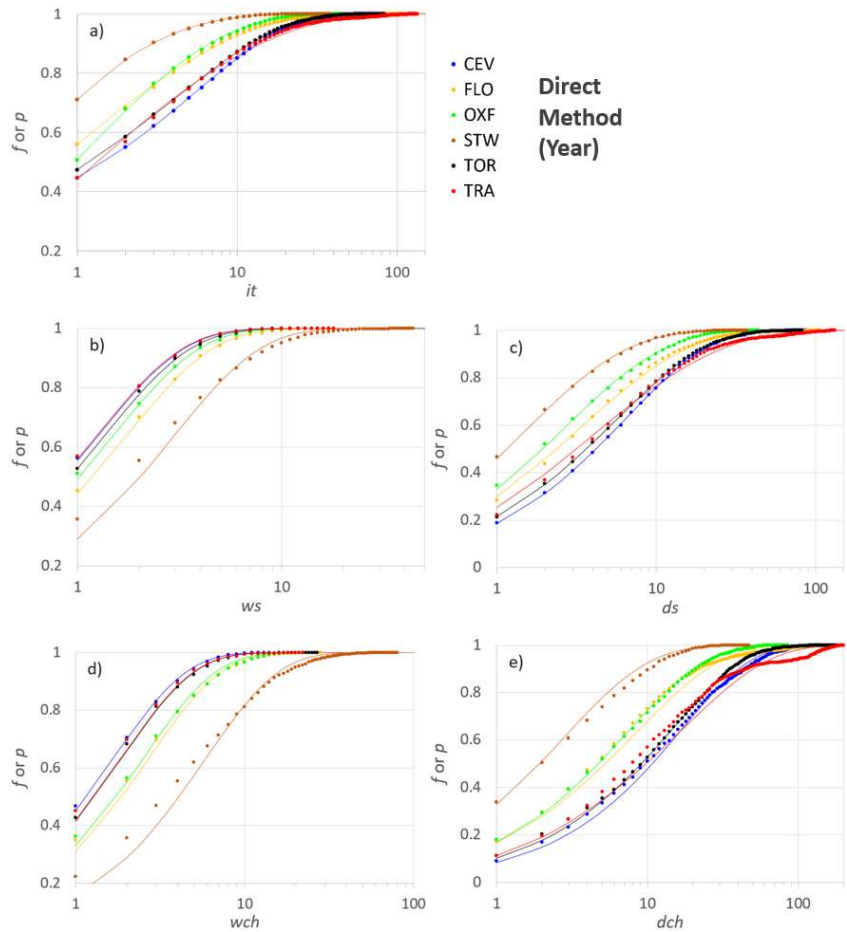

**Figure 8: Observed frequencies and fitted probability distributions for the six stations according to the direct method (DM) and for the period "Year". The variables on the x-axis are in logarithmic scale.**

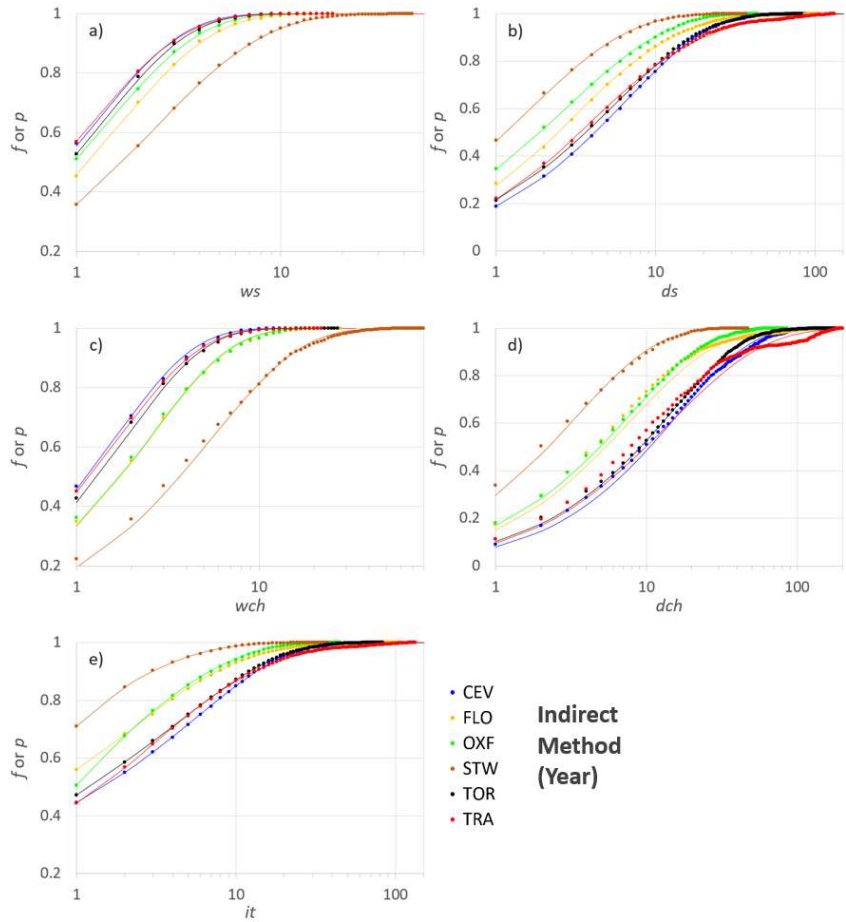

**Figure 9: Observed frequencies and fitted probability distributions for the six stations according to the indirect method (IM) and for the period "Year". The variables on the x-axis are in logarithmic scale.**

Since the previous results show that the main difference between the two methods concerns the ability to model *ws* and *wch*, and to a lesser extent *dch*, the plots in **Fig. 10** depict the difference in the cumulated frequencies for these variables modelled through the two methods for S1 (panels a, c, e) and S2 (panels b, d, f), respectively. These results further emphasize how the IM improves the performances in most of the cases, given the high fraction of points located in the portion of the plots below the 1:1 line (i.e., differences in the DM are larger than in the IM). Overall, most of the differences lies along the identity line, suggesting a good performance for both methods, but with large discrepancies observed for STW, OXF and FLO, thus for the cases in which the renewal property needs to be relaxed.

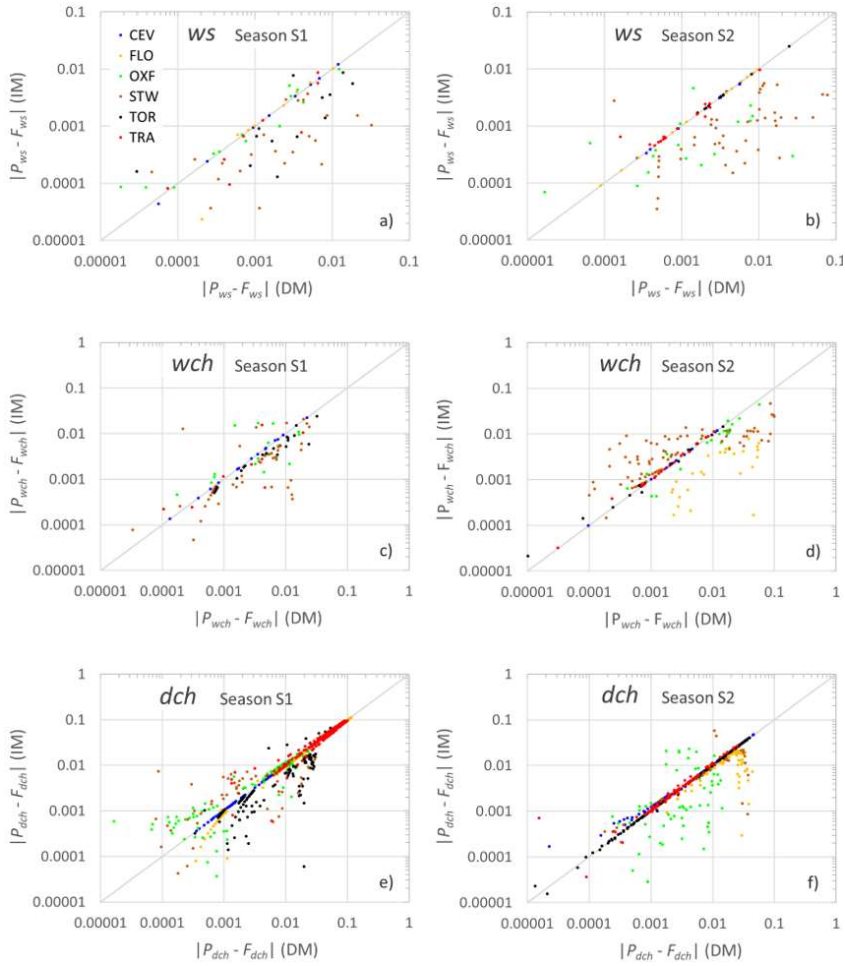

**Figure 10**: Scatterplots of the absolute difference between observed frequencies and fitted probabilities with the DM (x-axis) against the IM (y-axis) for *ws* (a, b), *wch* (c, d), and *dch* (e, f).

The Lerch family distribution also allows predicting the probability of extremes of the time variables. The overall consistence of the latter can be observed in **Fig. 11**, where the empirical 99$^{th}$ percentiles, $Q_{0.99}$, are compared with the estimated ones, for all stations and periods, when applying the DM (**Fig. 11a**) and the IM (**Fig. 11b**). **Figure 11a** shows that the dots are quite close to the line of perfect agreement, with only a few exceptions likely due to the limited sample size, as for the Sicilian stations for season S1. A slight improvement can be obtained by applying the IM, as shown by the decrease in the standard error of estimate (SEE) reported in the figures. These results suggest that the poor performances observed for the DM is some cases does not significantly affect the results on the tails, but only the accuracy on the most frequent data.

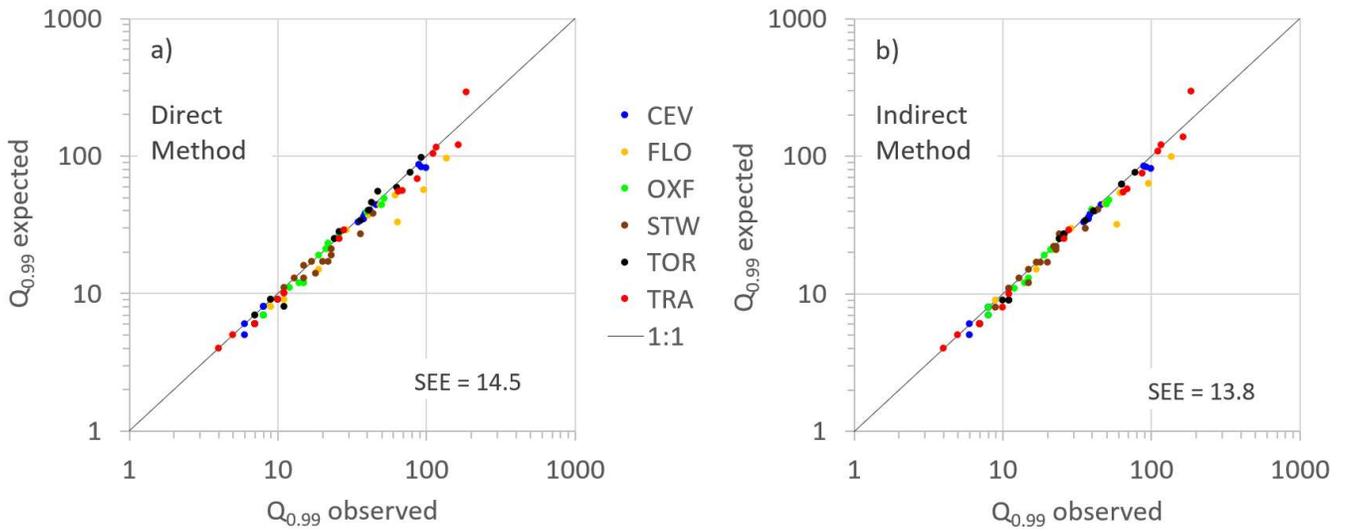

Figure 11: Comparison between empirical and theoretical quantile $Q_{0.99}$, calculated according to the DM (a) and the IM (b) for all stations, periods, and time variables bundled together.

## 5 Discussion

The fitting of the Lerch distribution to the selected six stations extends the studies previously carried out for the stations in Sicily and Piedmont (Agnese at al., 2014; Baiamonte et al., 2019). The adequacy of this distribution in fitting *it* is confirmed when data recorded over OXF and STW are considered, despite the rather different rainfall patterns of these latter stations compared with those previously analysed. Most notably, the observed frequencies for *it* are already well reproduced at the annual scale, stressing how splitting the datasets into sub-periods is not strictly necessary to correctly reproduce the probabilistic law of *it* in any of the rainfall regimes under consideration. The scientific literature on the statistical inference of rainfall inter-arrival times is still rather sparse, so further evidence of the suitability of the Lerch family to reproduce *it* distributions in a wide range of rainfall regimes is encouraging for further applications of the methodology.

However, the results obtained for *ws* and *ds* with the DM highlight how a good fit of *it* does not necessarily guarantee a satisfactory reproduction of the frequency of these derived quantities and, in particular, how the geometric distribution is not always adequate in describing wet spells. The ability to better reproduce the observed frequencies of *ws* with the IM, thus with distributions relaxing the memoryless property, suggests the presence of an inner structure in multiday rainfall events, which manifests in a not-constant rain probability within the event itself. Usually, the internal structure observed in sub-daily (e.g., 10-min) rainfall records is assumed to vanish at the daily aggregation steps (e.g., Ridolfi et al., 2011), but the results here reported may contradict this assumption in some of the investigated sites.

The need to resort to a more complex distribution than the geometric to reproduce the probabilistic structure of *ws* has been highlighted in the literature, especially for long spells, by the studies of Berger and Goossens (1983) for rainfall data in Belgium and Deni and Jemain (2010) in Malaysia. The inadequacy of the geometric distribution is here confirmed for long *ws* (> 10 days) with relatively high frequency, such as the case of STW where the DM returned poor performances. It is important to underline how the unsatisfactory fitting of a geometric distribution may not necessarily translate in the presence of memory in the rainfall series. In this regard, on the one hand, the results observed for the stations of TRA and FLO may suggest other reasons behind the poor performance of the geometric distribution in modelling *ws* as deducted from *it*. The geometric distribution seems to be a suitable choice when the data are analysed separately for the two seasons, which might imply that the complex structure of the *ws* distribution observed over the entire year is not associated with an actual relaxing of the renewal property but with a mixing of *ws* samples collected during two rather distinct seasons.

On the other hand, the geometric distribution seems to perform poorly on STW regardless of seasonality, which is actually quite limited for this station. The STW station seems to represent a case where the memoryless property is violated, as also confirmed by the inspection of the $\frac{S_{r+1}}{S_r}$ ratios.

Splitting the entire dataset into sub-periods seems to improve the performance of fittings crosswise, for both the DM and the IM. This result is well suited for possible implementations of the methodology for operational applications related to ecohydrology models (e.g., D'Odorico et al., 2000; Petrie and Brunsell, 2011) and stochastic weather generators (e.g., Paek et al., 2023). In these fields, to express the climatic component of weather variables (Semenov et al., 1998), not only the overall probabilistic structure of rainfall needs to be reproduced but also information on a seasonal or even sub-seasonal (i.e., monthly) scale is required. Other studies, carried out over regions characterized by a climate with no distinct monsoon seasons, have also highlighted the importance of focusing on either the dry summer seasons or wet winter seasons (Caloiero and Coscarelli, 2020; Paton, 2022; Raymond et al., 2016). Wan et al. (2015), also suggesting the need to account for the seasonality to properly reproduce the duration of *ws* for Canada using a Markov chain method.

Another consequence of the inadequacy of the geometric distribution in describing wet periods is that the daily structure of rainfall needs to be taken into account for modelling processes such as seasonal dynamics of soil moisture and vegetation. Ratan and Venugopal (2013) did an assessment for tropical areas, using satellite rainfall data. They found wet spell durations with a peak at one day for dry regions, while the duration of 2-4 days is predominant for humid areas. A similar but reversed observation was made for dry spells, resulting in one day for humid areas, 3-4 days for dry areas.

For some cases, the results obtained for the IM suggest that the classical application of the geometric distribution for *ws*, and of the polylog series for *ds*, provides satisfactory modelling of the observed frequencies. However, there is no substantial benefits in using these two distributions over using a 3-parameter Lerch distribution. Indeed, a similar number of fitting parameters is required in DM and IM, but with the obvious drawback of having an increasing computational cost due to the need to perform two independent fittings (for *ws* and *ds*) compared to a single one (*it* only). In this context, peculiar is the case of STW, where the geometric distribution is never selected for *ws* and a 2-parameter distribution is always required, although a 1-parameter distribution is used in this site for *ds*. This circumstance suggests that a reliable fit of the two quantities can still be achieved without increasing the total number of parameters when compared with the 3-parameter Lerch.

Finally, it is worth mentioning that the models proposed in this paper are local, hence spatial dependency in parameters may need to be accounted for applications to multiple stations located at shorter distance.

# 6 Conclusions

In this paper, daily rainfall data belonging to a large range of rainfall regimes across Europe (latitudes 38° - 58° N) have been analysed to model the frequency distribution of some key rainfall time variables. By using two different methods, the assumption of the renewal property that implies the geometric distribution of wet spells has been investigated. First, a direct method (DM), where the geometric distribution of wet spells is assumed, has been applied. Second, the latter assumption is relaxed by using an indirect method (IM) where wet spells and dry spells have been modelled separately, hence including the possibility of accounting for a not constant rain probability inside the rainfall cluster.

As a general rule, the results of comparing the DM and the IM suggest that the Lerch distribution can be successfully used for both inter-arrival times and dry spells in a wide variety of rainfall regimes, whereas a preliminary analysis of the memory property (e.g., on the $\frac{S_{r+1}}{S_r}$ ratios) may be needed to assess the reliability of the wet spells derived from the inter-arrival times modelled with the DM using the Lerch family. When signals of memory are detected, the IM is recommended, as it is better suited for a wider range of conditions albeit with a potentially larger number of fitting parameters.

The analysis was extended to include two additional time variables, strongly associated with wet and dry spells, referred to as wet and dry chains. These variables extend the concept of wet and dry spells to sequences characterized by an interruption of one no-rainy or one rainy day, respectively, as they represent two quantities that may be of interest for practical hydrological applications. The results obtained for the two chains generally reflect the findings obtained for the spells, albeit highlighting additional difficulties in the probabilistic modelling, especially at sites where the sample size may become a limiting factor.

The effects of the seasonality on the results were also addressed, splitting the data in two 6-month sub-periods. This separation tends to improve the performances for both the DM and the IM, stressing how in most of the sites the DM applied to seasonal data is still a suitable straightforward approach. The results of this study may help in scenario simulations of drought and flood events, considering that probabilistic functions, such as those applied in this work, are at the base of stochastic climate modelling.

Future research aimed at investigating the neighbouring location effects on parameter values will be developed.

**Competing Interests.** The contact author has declared that neither of the authors has any competing interests.

**Financial support.** This research did not receive any specific grant from funding agencies in the public, commercial, or not-for-profit sectors.

**Acknowledgements.** The Authors thank Prof. Nicholas Howden, School of Civil, Aerospace and Mechanical Engineering, University of Bristol, for providing the Oxford rainfall data. The authors would like to thank the anonymous reviewers and the editor for their valuable comments.

# References

Agnese, C., Baiamonte, G., and Cammalleri, C.: Modelling the occurrence of rainy days under a typical Mediterranean climate, Adv. Water Resour., 64: 62–76, 2014.

Baiamonte, G., Mercalli, L., Cat Berro, D., Agnese, C., and Ferraris, S.: Modelling the frequency distribution of inter-arrival times from daily precipitation time-series in North-West Italy, Hydrol. Res., 50(1): 339–357, 2019.

Berger, A., and Goossens, C.: Persistence of wet and dry spells at Uccle (Belgium), J. Climatol., 3(1): 21-34, 1983.

Bonsal, B.R., and Lawford, R.G.: Teleconnections between El Niño and La Niña events and summer extended dry spells on the Canadian prairies, Int. J. Climatol., 19(13): 1445–1458, 1999.

Caloiero, T., and Coscarelli, R.: Analysis of the characteristics of dry and wet spells in a Mediterranean region, Environ. Process., 7: 691-701 doi:10.1007/s40710-020-00454-3, 2020.

Chatfield, C.: Wet and dry spells. Weather, 21(9): 308–310, 1966.

Chowdhury, R.K., and Beecham, S.: Characterization of rainfall spells for urban water management, Int. J. Climatol., 33: 959–967, 2013.

Deni, M.S., and Jemain, A.A.: Mixed log series geometric distribution for sequences of dry days. Atmospheric Research, 92: 236–243, 2009.

Deni, M.S., Jemain, A.A., and Ibrahim, K.: The best probability models for dry and wet spells in Peninsular Malaysia during monsoon seasons, Int. J. Climatol., 30: 1194-1205, doi:10.1002/joc.1972, 2010.

Dobi-Wantuch, I., Mika, J., and Szeidl, L.: Modelling wet and dry spells with mixture distributions, Meteorol. Atm. Phys., 73: 245-256 doi:10.1007/s007030050076, 2000.

D'Odorico, P., Ridolfi, L., Porporato, A., and Rodriguez-Iturbe, I.: Preferential states of seasonal soil moisture: The impact of climate fluctuations, Water Resour. Res., 36(8): 2209-2219 doi:10.1029/2000WR900103, 2000.

Dey, P.: On the structure of the intermittency of rainfall, Water Res. Manag., 37:1461-1472, doi:10.1007/s11269-023-03441-z, 2023.

El Hafyani, M., and El Himdi, K.: A Comparative Study of Geometric and Exponential Laws in Modelling the Distribution of Daily Precipitation Durations, IOP Conf. Ser.: Earth Environ. Sci. Volume 1006, 12th International Conference on Environmental Science and Technology, 1006 012005, 2022.

FUME package, Santander Meteorology Group. R package version 1.0, https://CRAN.R-project.org/package=fume, 2012.

Gilbert, R.O.: Statistical methods for environmental pollution monitoring, Van Nostrand Reinhold Company, 334 pp, 1987.

Gocic, M., and Trajkovic, S.: Analysis of changes in meteorological variables using Mann-Kendall and Sen's slope estimator statistical tests in Serbia, Global Planet. Change, 1001:72–182, 2013.

Green, J. R.: A generalized probability model for sequences of wet and dry day, Monthly Weather Review, 98(3):238-241, 1970.

Gupta, P.L., Gupta, R.C., Ong, S.-H., and Srivastava, H.: A class of Hurwitz–Lerch-Zeta distributions and their applications in reliability, Appl. Mathem. Comput., 196(2): 521–531, 2008.

Hamed, K.H., and Rao, A.R.: A modified Mann-Kendall trend test for autocorrelated data, J. Hydrol., 204(1):182–196, 1998.

Hershfield, D.M.: A comparison of conditional and unconditional probabilities for wet- and dry-day sequences, J. Appl. Meteorol., 9(5): 825-827, 1970.

Hope, A.C.: A simplified Montecarlo significance test procedure, J. Royal Stat. Soc. Series B (Methodological), 30(3):582–598, 1968.

Hughes, J.P., and Guttorp, P: Incorporating spatial dependence and atmospheric data in a model of precipitation, J. Appl. Meteorol. 33(12): 1503–1515. https://doi.org/10.1175/1520-0450(1994)033<1503:ISDAAD>2.0.CO;2.

Kottegoda, N.T., Rosso, R.: Statistics, Probability, and Reliability for Civil and Environmental Engineers, McGraw-Hill, New York, NY, USA, 735 pp., 1997.

Martínez-Rodríguez, A.M., Sáez-Castillo, A.J., and Conde-Sánchez, A.: Modelling using an extended Yule distribution, Comput. Stat. Data Anal., 55(1): 863-873. doi:10.1016/j.csda.2010.07.014, 2011.

Mann, H.B.: Nonparametric tests against trend, Econometrica, 13(3):245–259, 1945.

Mhanna, M., and Bauwens, W.: A stochastic space-time model for the generation of daily rainfall in the Gaza Strip, Int. J. Climatol. 32: 1098–1112, 2012.

Osei, M.A., Amekudzi, L.K., and Quansah, E.: Characterisation of wet and dry spells and associated atmospheric dynamics at the Pra River catchment of Ghana, West Africa, J. Hydrol. Regional Studies, 34:100801. doi:10.1016/j.ejrh.2021.100801, 2021.

Paek, J., Pollanen, M., and Abdella, K.: A stochastic weather model for drought derivatives in arid regions: A case study in Qatar, Mathematics, 11(7): 1628 doi:10.3390/math11071628, 2023.

Paton, E.: Intermittency analysis of dry spell magnitude and timing using different spell definitions, J. Hydrol., 608:127645. doi:10.1016/j.jhydrol.2022.127645, 2022.

Petrie, M.D., and Brunsell, N.A. The role of precipitation variability on the ecohydrology of grasslands, Ecohydrol., 5(3):337-345 doi: 10.1002/eco.224, 2011.

Racsko, P., Szeidl, L., and Semenov, M.: A serial approach to local stochastic weather models, Ecol. Modell.: 57, 27-41, 1991.

Ratan, R., and Venugopal, V.: 2013. Wet and dry spell characteristics of global tropical rainfall, Water Resour. Res., 49(6):3830–3841, 2013.

Raymond, F., Ullmann, A., Camberlin, P., Drobinski, P., and Chateau Smith, C.: Extreme dry spell detection and climatology over the Mediterranean Basin during the wet season, Geophys. Res. Lett., 43: 7196–7204 doi:10.1002/2016GL069758, 2016.

Robertson, A.W., Kirshner, S., and Smyth, P.: Downscaling of daily rainfall occurrence over Northeast Brazil using a hidden Markov model, J. Climate 17:4407–4424, 2004.

Robertson, A.W., Kirshner, S., Smyth, P., Charles, S.P., Smyth Bates, B.C.: Subseasonal-to-interdecadal variability of the Australian monsoon over North Queensland, Q. J. Roy. Meteor. Soc. 132: 519–542, 2006.

Ridolfi, L., D'Odorico, P., and Laio, F.: Noise-Induced Phenomena in the Environmental Sciences, Cambridge University Press, 313 pp., doi:10.1017/CBO9780511984730, 2011.

Schleiss, M., and Smith, J.A.: Two simple metrics for quantifying rainfall intermittency: The burstiness and memory of interamount times, J. Hydrometeorol., 17(1):421-436 doi:10.1175/JHM-D-15-0078.1, 2016.

Semenov, M.A., Brooks, R.J., Barrow, E.M., and Richardson, C.W.: Comparison of the WGEN and LARS-WG stochastic weather generators for diverse climates, Clim. Res., 10:95-107 https://www.jstor.org/stable/24865958, 1998.

Wan, H., Zhang, X., and Barrow, E.M.: Stochastic modelling of daily precipitation in Canada. Atmos. Ocean, 43(1):23-32, 2015.

Wilks, D.S.: Multisite generalization of a daily stochastic precipitation generation model, J. Hydrol. 210: 178–191, 1998.

Wilks, D.S.: Interannual variability and extreme-value characteristics of several stochastic daily precipitation models, Agr. Forest Meteorol., 93(3):153-169 doi:10.1016/S0168-1923(98)00125-7.J, 1999.

Wilks, S.S.: The Large-Sample Distribution of the Likelihood Ratio for Testing Composite Hypotheses, Ann. Math. Statist., 9(1): 60-62 http://dx.doi.org/10.1214/aoms/1177732360, 1938.

Zhang, J., Li, L., Wu, Z., and Li, X.: Prolonged dry spells in recent decades over north-Central China and their association with a northward shift in planetary waves, Int. J. Climatol., 35(15):4829–4842 doi:10.1002/joc.4337, 2015.

Zolina, O., Simmer, C., Belyaev, K., Gulev, S.K., and Koltermann, P.: Changes in the duration of European wet and dry spells during the last 60 years, J. Climate, 26(6): 2022-2047, doi:10.1175/JCLI-D-11-00498.1, 2013.

Zörnig, P., and Altmann, G.: Unified representation of Zipf distributions, Computational Statistics & Data Analysis, 19(4): 461-473, 1995.